\documentclass[a4,12pt]{amsart}

\usepackage[english]{babel}
\usepackage[T1]{fontenc}

\usepackage{hyperref}

\usepackage[utf8]{inputenc}
\usepackage{amsmath,amssymb,mathrsfs}
\usepackage{amscd}
\usepackage{tikz}

\numberwithin{equation}{section}
\newtheorem{theo}{Theorem}[section]

\newtheorem{prop}{Proposition}[section]
\newtheorem{lem}{Lemma}[section]

\theoremstyle{remark} 

\numberwithin{figure}{section}
\theoremstyle{assumption}
\newtheorem*{assumption}{Assumption (A)}

\newcommand{\br}{{\mathbb{R}}}
\newcommand{\bn}{{\mathbb{N}}}
\newcommand{\bc}{{\mathbb{C}}}

\newcommand{\A}{\textbf{A}}
\newcommand{\W}{\textbf{\textup{W}}}
\newcommand{\w}{\textbf{\textup{w}}}
\newcommand{\T}{\mathcal{T}}

\newenvironment{prof}
	{\textbf{Proof.}}
	{\hfill $\square$\vskip 8pt}

\title[Distribution of magnetic eigenvalues]{Eigenvalues behaviours for self-adjoint Pauli 
operators with unsigned perturbations and admissible magnetic fields}

\author{Diomba \textsc{Sambou}}
\address{Departamento de Matem\'aticas, Facultad de Matem\'aticas, 
Pontificia Universidad Cat\'olica de Chile, Vicu\~na Mackenna 
4860, Santiago de Chile}
\email{disambou@mat.uc.cl}

\author{Amal \textsc{Taarabt}}
\address{Instituto de F\'isica, 
Pontificia Universidad Cat\'olica de Chile, Vicu\~na Mackenna 
4860, Santiago de Chile}
\email{ataarabt@fis.puc.cl}

\thanks{The two authors have been supported by the Chilean Program 
N\'ucleo Milenio de F\'isica Matem\'atica RC$120002$. D. Sambou is 
supported by the Chilean Fondecyt Grant $3170411$. The authors are grateful 
to J.-F. Bony for his suggestion in the use of the reduction \eqref{eq4,2}, 
and G. Raikov for his helpful suggestions during the revision of this note.
}

\keywords{Pauli operators, eigenvalues, sign-indefinite perturbations}
\subjclass[2010]{Primary: 35B34; Secondary: 35P25.}

\begin{document}

\begin{abstract}
We investigate the discrete spectrum behaviour for the 2d Pauli 
operator with nonconstant magnetic field, perturbed by a \textit{sign-indefinite} 
self-adjoint electric potential which decays polynomially at infinity. A localisation 
of the eigenvalues and new asymptotics are established. 
\end{abstract}

\maketitle

\section{Introduction and results}
We consider a quantum spin-$\tfrac12$ non-relativistic particle submitted to 
an electromagnetic field and described by the Pauli operator 
\begin{equation}\label{Pauli H}
H(b,V) := \begin{pmatrix}
   (-i\nabla - \A)^{2} - b & 0 \\
   0 & (-i\nabla - \A)^{2} + b
\end{pmatrix} + V \quad \ \mathrm{on} \quad \mathrm{L}^2(\br^2,\bc^2),
\end{equation}
where $V = V(x)$, $x \in \br^2$, is a $2\times 2$ Hermitian matrix-valued 
potential, and $\A$ is a vector potential generating the magnetic field 
$b = \nabla \wedge \A$. We assume $b = b(x)$ to be an admissible magnetic 
field in the sense there exists a constant $b_0 > 0$ such that 
\begin{equation}
b(x) = b_0 +\tilde b(x),
\end{equation} 
with the Poisson equation $\Delta \tilde \varphi 
= \tilde b$ admitting a solution $\tilde\varphi\in \mathrm{C}^2(\br^2)$ 
which satisfies $\sup_{x\in\br^2}|D^\alpha\tilde\varphi(x)| <\infty$
for $\alpha\in\bn$, $|\alpha|\le2$. We refer for instance to \cite{ra}
for examples of admissible magnetic fields.

\medskip

In the unperturbed case where $V = 0$, the spectrum of $H(b,0)$ belongs to 
$\lbrace 0 \rbrace \cup [\zeta,+\infty)$ with 
$\zeta = 2b_0 e^{-2\mathrm{osc}(\tilde\varphi)}$ and $\mathrm{osc}(\tilde\varphi) :=\sup_{x\in\br^2}\tilde\varphi(x)-\inf_{x\in\br^2}\tilde\varphi(x)$.
Furthermore, $0$ is an eigenvalue of infinite multiplicity (see e.g. \cite{ra}). 
Notice that in the constant magnetic field case $b = b_0$, we have 
$\zeta = 2b_0$ the first Landau level of the shifted Schrödinger operator 
$(-i\nabla - \A)^{2} + b$. The case where $V$ is of definite sign has 
been already studied in \cite{ra}. In the present note, we are 
interested in the sign-indefinite potentials $V$ of the form
\begin{equation}\label{eq1,11}
V(x) := \begin{pmatrix}
   0 & \overline{U(x)} \\
   U(x) & 0
\end{pmatrix}, \quad \mathrm{for}\quad x\in\br^2,
\end{equation}
where the function $U(x) \in \bc$ satisfies 
\begin{equation}\label{eq1,13}
\vert U(x) \vert= \mathcal{O} (\langle x \rangle^{-m}),
\quad \langle x \rangle : = \sqrt{1 + \vert x 
\vert^{2}}, \quad\mathrm{for\ some}\quad m > 0.
\end{equation}

\noindent
\textbf{Remark.}
The potentials $V$ of the form \eqref{eq1,11} are 
\textit{sign-indefinite} since their eigenvalues are given by 
$\pm{\vert U(x) \vert}$.

\medskip

Under condition \eqref{eq1,13}, $V$ is relatively compact with respect 
to the operator $H(b,0)$ so that $\sigma_{\textup{ess}} \big( H(b,V) \big) = 
\sigma_{\textup{ess}} \big( H(b,0) \big)$, where $\sigma_\mathrm{ess}$ denotes 
the essential spectrum. However, $H(b,V)$ may have a discrete spectrum 
$\sigma_{\textup{disc}} \big( H(b,V) \big)$ that can accumulate at $0$. 
The aim of this note is to study this discrete spectrum near the low ground 
energy $0$. The novelty of this work arises from sign-indefinite perturbations 
we consider and behaviours we obtain. This is probably one of the first works 
dealing with sign-indefinite perturbations in a magnetic framework, see also 
the recent work \cite{diom} where the case of 3d Pauli operators 
are studied in a resonance point of view. 
We denote 
\begin{equation}\label{h1}
H_\pm := (-i\nabla - \textbf{A})^{2} \pm b \quad \mathrm{on} \quad 
\mathrm{L}^{2}(\br^2) := \mathrm{L}^2(\br^2,\bc),
\end{equation}
the component operators of the Pauli operator \eqref{Pauli H}. Let $p := p(b)$ be 
the orthogonal projection of $\mathrm{L}^{2}(\br^2)$ onto the (infinite dimensional) 
kernel of $H_-$. The corresponding projection in the constant magnetic field case 
will be denoted $p_0 := p(b_0)$. For a bounded operator $B \in \mathscr{L} 
\big(\mathrm{L}^{2}(\br^2) \big)$, we introduce the operator $\W(B)$  defined by
\begin{equation}\label{eq2,2}
\big(\W(B) f \big) (x) := 
\overline{U} (x) B (U f) (x).
\end{equation}
If $I$ denotes the identity operator on $\mathrm{L}^{2}(\br^2)$, then
$\W(I)$ is the multiplication operator by the function $x \longmapsto \vert U(x) 
\vert^{2}$. This function will be denoted $\W(I)$ again. Our results are strongly 
related to the operator $\W (B)$ through the Toepliz operator 
\begin{equation}\label{Toepliz}
p\W(B)p, \qquad B = I \quad \textup{or} \quad H_+^{-1}.
\end{equation}
Since the spectrum of the invertible operator $H_+$ belongs to $[\zeta,+\infty)$
and $U$ fulfils \eqref{eq1,13}, then it follows from \cite[Lemma 3.5]{ra} that 
the positive self-adjoint operators $p \W(I) p$ and $p \W \big( H_+^{-1} \big) p$ 
are compact on $\mathrm{L}^{2}(\br^2)$. For further use, let us introduce the following:

\noindent
\begin{assumption}\label{A}
The function $U \in C^1(\br^2)$ satisfies
\begin{equation}
0 \leq U(x) \leq C \langle x \rangle^{-m}, \hspace{0.2cm} 
\vert \nabla U(x) \vert \leq C \langle x \rangle^{-m-1}, \quad x \in \br^2,
\end{equation}
for some constants $C > 0$, $m>0$, and
$U(x) = U_0 \big( \tfrac{x}{|x|} \big) |x|^{-m}(1+o(1)), \ |x|\to +\infty$, with
 $0 \not\equiv U_0\in C^0(\mathbb{S}^1)$.
\end{assumption}

\noindent
\textbf{Integrated density of states (IDS):} 
For $x \in \mathbb{R}^{2}$, let 
$\chi_{T,x}$ be the characteristic 
function of the square $x + (T/2,T/2)^{2}$ with 
$T > 0$. Denote ${\bf 1}_{I}(H_-)$ the spectral 
projection of $H_-$ on the interval 
$I \subset \mathbb{R}$.
A non-increasing 
function $g : \mathbb{R} \longrightarrow 
[0,+\infty)$ is called an IDS for the operator
$H_-$ if it satisfies for any 
$x \in \mathbb{R}^{2}$
$$
g(t) = \lim_{T\rightarrow\infty} T^{-2} 
\hspace{0.5mm} \textup{Tr} 
\hspace{0.5mm} \big[ \chi_{T,x} 
{\bf 1}_{(-\infty,t)}(H_-) 
\chi_{T,x} \big],
$$ for each point $t$ of continuity of $g$ (see e.g. \cite{ra}). 

\medskip

\noindent
\textbf{Remark.}
If $b = b_0$ is constant, then there exists naturally an IDS for the operator 
$H_-$ given by 
$$
g(t) = \frac{b_{0}}{2\pi} 
\sum_{q=0}^{\infty} \chi_{\mathbb{R}_{+}} 
(t - 2 b_{0} q), \quad t \in \mathbb{R}, \quad 
\chi_{\mathbb{R}_{+}} (t) = 
\begin{cases} 
1 & \text{if } t \in \br^+, \\ 
0 & \text{otherwise}. 
\end{cases}
$$ 
\noindent
In the next results, the discrete eigenvalues of the operator 
$H(b,e V)$ are counted according to their
multiplicity defined by \eqref{multiplicity}, for which we conjecture that it
coincides with the geometric multiplicity.

\begin{theo}\label{theo1} 
Assume that $V$ and $U$ fulfil \eqref{eq1,11} and \eqref{eq1,13} respectively. 
Then, there exists a discrete set $\mathcal{E} \subset \mathbb{R}$ such 
that for any $e \in \mathbb{R} \setminus \mathcal{E}$ and any $0 < r_0 \ll 1$, 
the following holds:
\begin{enumerate}
\item[(i)] Localization: If $z$ is a discrete eigenvalue of $H(b,e V)$
near zero, then $z \leq 0$.
\item[(ii)] Asymptotic: Suppose that 
$\# \big\lbrace z \in \sigma(p \W ( H_+^{-1})p) 
: z \geq r \big\rbrace \rightarrow + \infty$
as $r \searrow 0$. Then, there exists a positive sequence $(r_{\ell})_{\ell}$ tending 
to $0$ such that as $\ell \longrightarrow \infty$,
\begin{align}\label{eq2,5}
\# \big\lbrace z \in \sigma_{\mathrm{disc}} 
\left( H(b,e V) \right)& : -r_0e^2 \leq z 
< -r_{\ell}e^2 \big\rbrace\notag\\
&= \# \big\lbrace z \in \sigma( p \W (H_+^{-1}) 
p) : z \geq r_{\ell} \big\rbrace \big( 1 + o(1) \big).
\end{align}
\item[(iii)] Upper-bound: Assume that there exists an $IDS$ $g$ for  
$H_-$. If $\W(I)$ satisfies Assumption (A) and 
$$\textup{Tr} \hspace{0.6mm} 
\mathbf{1}_{[r(1+\nu),r(1-\nu)]} 
\big( p \W ( H_+^{-1}) p \big) = \textup{Tr} 
\mathbf{1}_{[r,1]} \big( p \W ( H_+^{-1}) p \big) \big( o(1) + \mathcal{O}(\nu) \big),$$ 
for any $0 < \nu \ll 1$, $r \searrow 0$, then, 
\begin{equation}\label{eq2,6}
 \limsup_{r \searrow 0} \frac{\# \big\lbrace z \in \sigma_{\mathrm{disc}} 
( H(b,e V) ) : -r_0 e^2 \leq z 
< -r e^2 \big\rbrace}{\# \big\lbrace z \in \sigma \left( \zeta^{-1} p \W(I) p \right) 
: z \geq r \big\rbrace} \leq 1.
\end{equation}
\end{enumerate}
\end{theo}

\noindent
Furthermore, if the magnetic field is constant ($i.e.$ $b = b_0$), we 
obtain the following

\begin{theo}\label{theo2} 
\mbox

\begin{itemize}
\item[(i)] Assume that
$\# \big\lbrace z \in \sigma\left(( p_{0}Up_{0} 
\big)^{\ast} p_{0}Up_{0} \right) : z \geq 2 r b_0 \big\rbrace =
\phi(r) \big( 1 + o(1) \big)$, $r \searrow 0$,
where $\phi \big( r(1 \pm \nu) \big) = \phi (r) \big( 1 + o(1) + 
\mathcal{O}(\nu) \big)$ for any $0 < \nu \ll 1$. Suppose moreover that 
$\textup{Tr} \hspace{0.6mm} \mathbf{1}_{[r,1]} 
\big( p_0 \W ( H_+^{-1}) p_0 \big) = \phi(r) \big( 1 + o(1) \big)$, 
$\phi(r) \longrightarrow + \infty$, $r \searrow 0$. Then, as $r \searrow 0$,
\begin{equation}\label{eq2.10}
\begin{split}
\# \big\lbrace z \in &
\sigma_\mathrm{disc} 
( H(b_0,e V)) : -r_0 e^2 \leq z 
< -r e^2 \big\rbrace \\
& = \# \big\lbrace z \in \sigma\left( \big( p_{0}Up_{0} 
\big)^{\ast} p_{0}Up_{0} \right) : z \geq 
2 r b_0 \big\rbrace \big( 1 + o(1) \big).
\end{split}
\end{equation}

\item[(ii)] Assume that $U$ satisfies Assumption (A). Then,
as $r \searrow 0$,
\begin{equation}\label{eq2100}
\begin{split}
\# \big\lbrace z \in & \sigma_{\mathrm{disc}} 
( H(b_0,e V)) : -r_0 e^2 \leq z 
< -r e^2 \big\rbrace \\
&= \# \big\lbrace z \in \sigma( p_{0}Up_{0}) : z \geq 
\big( 2 r b_0 \big)^\frac{1}{2} \big\rbrace \big( 1 + o(1) \big).
\end{split}
\end{equation}
\end{itemize}
\end{theo}

\noindent
\textbf{Remarks.}
\begin{enumerate}

\item[(i)] The proof Theorem \ref{theo2}, (ii) shows that 
$\# \big\lbrace z \in \sigma(p_0 \W ( H_+^{-1})p_0) 
: z \geq r \big\rbrace \rightarrow + \infty$
as $r \searrow 0$. Then, by Theorem \ref{theo1}, (ii), the asymptotic
\eqref{eq2,5} holds with $p = p_0$.

\item[(ii)] Notice that thanks to the asymptotics of \cite[Lemma 3.3]{rage}, 
\eqref{eq2100} implies that the number of eigenvalues of the operator 
$H(b_0,e V)$ near $0$ satisfies
\begin{equation}\label{eq2.11}
\# \big\lbrace z \in \sigma_{\textup{disc}} 
\big( H(b_0,e V) \big) : -r_0 e^2 \leq z 
< -r e^2 \big\rbrace 
= C_{m} \left( \frac{1}{2b_0} \right)^{1/m} r^{-1/m} \big( 1 + o(1) \big), 
\end{equation}
as $r \searrow 0$, where 
\begin{equation}\label{eq2.11+}
C_{m} := \frac{b_{0}}{4\pi} \int_{\mathbb{S}^{1}} U_{0}(t)^{2/m} dt.
\end{equation}
In particular, it holds from \eqref{eq2.11} that the eigenvalues of 
$H(b_0,e V)$ less than $-re^2$ accumulate at zero with an
accumulation rate of order $r^{-1/m}$, whereas it was of order 
$r^{-2/m}$ for $V$ of definite sign in \cite{ra}.

\item[(iii)] Otherwise, we can expect that this kind of accumulation 
also occurs near all the Landau levels $2 b_0 q$, $q\in\mathbb{N}$, 
of the operator $H(b_0,V)$. However, the spectral analysis is more 
difficult due to the contribution of the half-Pauli operators $H_\pm$
near each Landau level $2 b_0 q$, $q\in\mathbb{N}^\ast$.
\end{enumerate}


\section{Strategy of the Proofs}

We explain the main ideas of the proofs and the relationship between the 
initial operator and the new quantities we are going to introduce. First,
let us introduce some useful notations. For $\mathscr{H}$ a separable Hilbert 
space, we denote $S_{\infty}(\mathscr{H})$ (resp. $GL(\mathscr{H})$) the 
set of compact (resp. invertible) linear operators in $\mathscr{H}$.
Let $D \subseteq \mathbb{C}$ be a connected open set, $Z \subset D$ be a 
discrete and closed subset, $A : \overline{D} \backslash Z \longrightarrow 
GL(\mathscr{H})$ be a finite meromorphic operator-valued function 
\big(see e.g. \cite{go} and \cite[Section 4]{goh}) and Fredholm at each point of 
$Z$. For an operator $A$ that does not vanish on $\gamma$ a positive oriented 
contour, the index of $A$ with respect to $\gamma$ is defined by 
\begin{equation}\label{ind}
\textup{Ind}_{\gamma} \hspace{0.5mm} A := 
\frac{1}{2i\pi} \textup{tr} \int_{\gamma} A'(z)A(z)^{-1} 
dz = \frac{1}{2i\pi} \textup{tr} \int_{\gamma} A(z)^{-1} 
A'(z) dz.
\end{equation}

\subsection{Reduction of the problem}

Let us consider the punctured disk
$D(0,\epsilon)^\ast := \big\lbrace z \in \bc : 0 < |z| < \epsilon \big\rbrace$ 
for $0 < \epsilon < \zeta$. 
For $z\in D(0,\epsilon)^\ast$ small enough, we have
\begin{align*}
 \big( H(b,V) - z \big) &
 \begin{pmatrix} I & 0 \\ 
 -(H_+-z)^{-1}U & (H_+-z)^{-1}
 \end{pmatrix} \\
&  =\begin{pmatrix}
  H_--z-\overline{U}(H_+-z)^{-1}U & \overline{U}(H_+-z)^{-1}
  \\ 0&I
 \end{pmatrix},
\end{align*}
so that the following characterisation holds:
\begin{equation}\label{eq4,2}
H(b,V) - z  \quad \text{is invertible} 
\hspace{0.1cm} \Leftrightarrow \hspace{0.1cm} 
H_- - z - \overline{U} 
(H_+ - z)^{-1} U  \quad \text{is invertible}.
\end{equation}
Thus, we reduce the study of the discrete eigenvalues of $H(b,V)$ near $z = 0$ 
to the analysis of the non-invertibility of the operator
$H_- - z  - \overline{U} (H_+ - z)^{-1} U$. It is not difficult to prove the
following lemma which gives a new representation of the operator 
$\overline{U}(H_+-z)^{-1}U$. 

\begin{lem}\label{lem5,4} 
For $z$ small enough, the operator $\overline{U} (H_+ - z)^{-1} U$ 
admits the representation
\begin{equation}\label{eq5,12}
\overline{U} (H_+ - z)^{-1} U = \w^{\ast} \left( I + M(z) \right)\w,
\end{equation}
where 
$\w := H_+^{-1/2} U$
and 
\begin{equation}\label{eq5,121}
z \longmapsto M(z) := 
z \sum_{k \geq 0} 
z^{k} H_+^{-k - 1}
\end{equation}
is analytic near $z = 0$.
\end{lem}

\noindent
Let $R_-(z)$ denote the resolvent of the operator $H_-$. We have the following

\begin{lem}\label{lem5,5} 
For $z$ small enough, the operator-valued function
\begin{equation*}
D(0,\epsilon)^{\ast} \ni z \longmapsto 
\T_V(z) := \big( I + M(z) \big) \w R_- (z) \w^{\ast},
\end{equation*}
is analytic with values in 
$S_{\infty} \left( L^{2}(\br^2) \right)$.
\end{lem}

\noindent
\begin{prof}
Since $M(z)$ and $R_-(z)$ are well defined and analytic for 
$z$ in $D(0,\epsilon)^{\ast}$, then the analyticity of $\T_V(z)$ follows.
The compactness holds from that of $U R_-(z) \overline{U}$, 
by combining the diamagnetic inequality and \cite[Theorem 2.13]{sim}.
\end{prof}

\noindent
We have the following Birman-Schwinger principle:

\begin{prop}\label{prop5,2} 
For $z_0$ near zero, the following assertions are equivalent: 

\begin{itemize}
\item[(i)] $z_0$ is a discrete eigenvalue
of $H(b,V)$,
\item[(ii)] $I - \T_V(z_0)$ is not invertible.
\end{itemize}
\end{prop}

\noindent
\begin{prof}
Set $\mathscr{R}(z) := \big( H_- - z - \overline{U} (H_+ - z)^{-1} U \big)^{-1}$.
Then, the proof follows directly from \eqref{eq4,2}, the fact that $\mathscr{R}(z)$ 
and $\big( I + M (z) \big) \w \mathscr{R}(z) \w^{\ast}$ have 
the same poles (the discrete eigenvalues $z$) near $0$, together with the identity 
\begin{equation}\label{id resolvent}
\Big( I - \big( I + M(z) \big) \w R_- (z) \w^{\ast} \Big) 
\Big( I + \big( I + M (z) \big) \w \mathscr{R}(z) \w^{\ast} \Big) = I.
\end{equation}
\end{prof}

\noindent
In Proposition \ref{prop5,2}, (ii), $z_0$ is said to be a characteristic 
value of the operator-valued function $I - \T_V(\cdot)$. Sometimes, by abuse
of language, we will say that $z_0$ is a characteristic value of the 
operator $I - \T_V(z)$. The multiplicity 
of a discrete eigenvalue $z_0$ is defined by 
\begin{equation}\label{multiplicity}
\textup{mult}(z_0) := \textup{Ind}_{\gamma} 
\hspace{0.5mm} \Big( I - \T_V(\cdot) 
\Big),
\end{equation}
where $\gamma$ is a small positively oriented contour containing $z_0$ as the 
unique discrete eigenvalue of $H(b,V)$ \big(see \eqref{ind}\big). We will 
denote $\mathcal{Z} \big( I - \T_{V}(\cdot) \big)$ the set of characteristic 
values of $I - \T_{V}(\cdot)$.

\subsection{Sketch of proof of Theorem \ref{theo1}} 

As preparation, we point out some facts. Since $p$ is the orthogonal 
projection onto $\mathrm{ker} \, H_-$ and $p^\bot := 1-p$, then we have
\begin{align}
R_-(z)
=(H_--z)^{-1} p+(H_--z)^{-1}p^\bot\notag
=-z^{-1}p +(H_--z)^{-1}p^\bot.
\end{align}
In particular, this implies that
\begin{equation}\label{T_v}
\T_V(z)=-\frac{1}{z}\ \w p\w^\ast - z^{-1} M(z) \w 
 p\w^\ast +\big( I+M(z) \big)\w R_-(z)p^\bot\w^\ast.
\end{equation}
In the first term of the r.h.s. of \eqref{T_v}, write the operator $\w p\w^\ast$ 
as $\w p\w^\ast = (p\w^\ast)^\ast(p\w^\ast)$. By the definition
of $\w$ in Lemma \ref{lem5,4}, we have $\w^\ast \w = \overline{U} H_+^{-1} U$. Since 
$\sigma(H_+)\subset[\zeta,\infty)$, then we have
\begin{equation}\label{eqm}
(p\w^\ast) (p\w^\ast)^\ast = p\w^\ast\w p=p\W(H_+^{-1})p\le \zeta^{-1}\ p\W(I)p,
\end{equation}
where $\W(\bullet)$ is the operator defined by \eqref{eq2,2}.
According to Proposition~\ref{prop5,2}, the discrete eigenvalues
$z$ of the operator $H(b,e V)$ near $0$ are the characteristic values 
of the operator $I-\T_{e V}(z)$. Let us set
$K_V(z) := \w p\w^\ast - z A(z)$, where 
\begin{equation}
A(z) := -z^{-1} M(z) \w p\w^\ast +\big( I+M(z) \big)\w R_-(z)p^\bot\w^\ast.
\end{equation}
Thus, we have 
\begin{equation}\label{e,supp}
I-\T_{e V}(z) = I + \frac{e^2}{z} K_V(z) = I - \frac{K_V^{(e)}(\lambda)}{\lambda},
\end{equation}
with the rescaling $\lambda = -z/e^2$ and the operator $K_V^{(e)}(\lambda)$ defined by
$K_V^{(e)}(\lambda) := K_V(-\lambda e^2)$, so that $K_V^{(e)}(0) = K_V(0) = \w p \w^\ast$. 
Moreover, $\big( K_{V}^{(e)} \big)'(\lambda) = -e^2 K_V'(-\lambda e^2)$ so that 
$\big( K_{V}^{(e)} \big)'(0) = -e^2 K_V'(0)$. Let $\Pi_{0}$ be the orthogonal projection 
onto ${\rm ker} \, K_V(0)$. The compactness of the operator $K_V'(0) \Pi_{0}$ implies 
the existence of a discrete set $\lbrace e_{n} \rbrace \subset \mathbb{R}$ finite or 
infinite such that the operator $I + e^2 K_V'(0) \Pi_{0}$ is invertible for each 
$e \in \mathcal{E} := \mathbb{R} \setminus \lbrace e_{n} \rbrace$. For 
$L \in S_{\infty} \left( L^{2} (\mathbb{R}^2) \right)$, we set
\begin{equation}\label{eq2,1}
n_{+}(r,L) := \text{rank} \hspace{0.6mm} 
{\bf 1}_{[r,\infty)}(L), \quad r > 0,
\end{equation}
where ${\bf 1}_{[r,\infty)}(L)$ is 
the spectral projection of $L$ on the 
interval $[r,\infty)$. We have 
\begin{equation}\label{eq6,9}
n_{+} \big( r,\textup{\textbf{w}} 
p \textup{\textbf{w}}^{\ast} \big) 
= n_{+} \big( r,p \textup{\textbf{w}}^{\ast} 
\textup{\textbf{w}}p \big) = 
n_{+} \Big( r,p \textbf{\textup{W}} \big( 
H_+^{-1} \big) p \Big), \quad r > 0.
\end{equation} 
Then, \eqref{eqm} 
implies that
\begin{equation}\label{eq6,11}
n_{+} \big( r,\textup{\textbf{w}} 
p \textup{\textbf{w}}^{\ast} \big) 
\leq n_{+} \left( r,\zeta^{-1} p \textbf{\textup{W}} (I) p 
\right), \quad r > 0.
\end{equation}

\noindent
We return now to the proof of Theorem \ref{theo1}.

\medskip

\noindent
(i)-(ii): The claim (i) follows immediately from \cite[Corollary 3.4]{bo} with 
$z$ replaced by $\lambda = -z/e^{2}$, thanks to \eqref{e,supp}. To deal with the 
claim (ii), introduce the sector
$$
\mathcal{C}_{\alpha}(a,a') := \lbrace x + iy \in 
\mathbb{C} : a \leq x \leq a', -\alpha x \leq y \leq 
\alpha x \rbrace
,$$
with $a > 0$ tending to $0$, $a' > 0$ fixed, and $\alpha > 0$. 
Proposition~\ref{prop5,2} together with \eqref{e,supp} show that $z$ is a discrete
eigenvalue near zero if and only if $\lambda$ is a characteristic value of
$I - \T_{e V}(-\lambda e^2)$. Moreover, the proof of (i) shows that for $-r_0 e^2 \leq z 
< -r e^2$, $0 < r_0 \ll 1$, the characteristic values $\lambda = -z/e^2$ are concentrated 
in the sector $\lambda \in \mathcal{C}_{\alpha}(r,r_{0})
\cap \br$, for any $\alpha > 0$. Hence, by setting 
$\mathcal{N} \big( \mathcal{C}_{\alpha}(r,r_{0}) \cap \br \big) := 
\# \big\lbrace \lambda \in \mathcal{Z} \big( I - \T_{e V}(-\lambda e^2) \big) : \lambda 
\in \mathcal{C}_{\alpha}(r,r_{0}) \cap \br \big\rbrace$,
one has 
\begin{equation}\label{eq6,13}
\begin{split}
\# \big\lbrace z \in \sigma_{\textup{disc}} 
\big( H(b,e V) \big) : -r_0 e^2 \leq z 
< -r e^2 \big\rbrace = \mathcal{N} \big( \mathcal{C}_{\alpha}(r,r_{0}) \cap \br \big) + 
\mathcal{O}(1),
\end{split}
\end{equation}
where $0 < r_0 \ll 1$. For an interval $\Lambda \subset \mathbb{R}^\ast$, let
\begin{equation}\label{eq-}
n(\Lambda) := \textup{Tr} \hspace{0.6mm} \mathbf{1}_{\Lambda} \big( 
K_V(0) \big),
\end{equation} 
be the number of eigenvalues of the operator 
$K_V(0)$ lying in $\Lambda$ and counted according to their 
multiplicity.
In view of \eqref{eq6,11}, we have
$
n \big( [r,r_0] \big) \leq n_{+} 
\left( r,\zeta^{-1}p \textbf{\textup{W}} (I) p
\right),
$
so that (ii) follows from \eqref{eq6,13} together with 
\cite[Corollary 3.9]{bo} and \eqref{eq6,9}.

\medskip

\noindent
(iii): Concerning (iii), if there exists an IDS for the operator $H_-$ and 
if the function $\textbf{\textup{W}} (I)$ satisfies Assumption (A), then by 
\cite[Lemma 3.3]{rage} we have 
$
n_{+} \left( r,
\zeta^{-1} p \textbf{\textup{W}} (I) p 
\right)
= \widetilde{C}_m (\zeta r)^{-1/m} 
\big( 1 + o(1) \big)$, $r \searrow 0$, for some constant $\widetilde{C}_m > 0$. 
Otherwise, \cite[Theorem 3.7]{bo} implies that for any $\nu > 0$ small enough, there
exists $r(\nu) > 0$ such that for all $0 < r < r(\nu)$, we have
\begin{align}\label{eq+}
\mathcal{N} \big( \mathcal{C}_{\alpha}(r,1) \cap \br \big) 
&=n \big( [r,1] \big) \big( 1 + \mathcal{O} \big( 
\nu \vert \ln \nu \vert^{2} \big) \big) \notag\\
&+ \mathcal{O} \big( \vert \ln \nu \vert^{2} 
\big) n \big( [r(1-\nu),r(1+\nu)] \big) + 
\mathcal{O}_{\nu}(1),
\end{align}
where the $\mathcal{O}$'s are uniform with respect 
to $r$, $\nu$ but the $\mathcal{O}_{\nu}$ may depend on $\nu$.
Since we have $n \big( [r,1] \big) \le 
n_{+} \left( r,\zeta^{-1} p \textbf{\textup{W}} (I) p \right)$, then if 
$$ \textup{Tr} \hspace{0.6mm} \mathbf{1}_{[r(1+\nu),r(1-\nu)]} 
\big( p \W ( H_+^{-1}) p \big) = \textup{Tr} \hspace{0.6mm} 
\mathbf{1}_{[r,1]} \big( p \W ( H_+^{-1}) p \big) \big( o(1) + O(\nu) \big),$$
we deduce from \eqref{eq+} that
\begin{equation}\label{eq++}
\begin{split}
\mathcal{N} \big( \mathcal{C}_{\alpha}(r,1) \cap \br \big) & \le
n_{+} \left( r, \zeta^{-1} p \textbf{\textup{W}} (I) p \right) \big( 1 + \mathcal{O} \big( 
\nu \vert \ln \nu \vert^{2} \big) \big) \\ 
& + \mathcal{O} \big( \vert \ln \nu \vert^{2} \big) n_{+} \left( r,
\zeta^{-1} p \textbf{\textup{W}} (I) p \right) \big( o(1) + \mathcal{O}(\nu) \big) + \mathcal{O}_{\nu}(1).
\end{split}
\end{equation}
Since $\mathcal{N} \big( \mathcal{C}_{\alpha}(r,r_{0}) \cap \br \big) = 
\mathcal{N} \big( \mathcal{C}_{\alpha}(r,1) \cap \br \big) + \mathcal{O}(1)$, then 
putting this together with \eqref{eq6,13} and \eqref{eq++}, we get
\begin{align*}
\limsup_{r \searrow 0}& \frac{\# \Big\lbrace z \in \sigma_{\mathrm{disc}} 
( H(b,e V) ) : -r_0e^2 \leq z 
< -r e^2 \Big\rbrace}{n_{+} \left( r,
\zeta^{-1} p \textbf{\textup{W}} (I) p \right)}\\
&\leq 1 + \mathcal{O} \big( 
\nu \vert \ln \nu \vert^{2} \big) + \mathcal{O} \big( \vert \ln \nu \vert^{2} \big) 
\mathcal{O}(\nu).
\end{align*}
Now, letting $\nu$ tend to $0$, the claim (iii) follows immediately.

\subsection{Sketch of Proof of Theorem \ref{theo2}}

If the magnetic field is constant, then 
$(2b_{0})^{-1} p_{0} = H_+^{-1} p_{0} \le H_+^{-1}$. This implies that
\begin{equation}\label{eq6.18}
(2b_{0})^{-1} \big( p_{0} U p_{0} \big)^{\ast} \big( p_{0} U p_{0} \big) 
\le p_{0} \textbf{\textup{W}} \big( H_+^{-1} 
\big) p_{0}.
\end{equation}

\medskip

\noindent
(i): If the assumptions of item (i) are satisfied, then we have
$$n \big( \big[ r(1 - \nu),r(1 + \nu) \big] \big) = 
n \big( [r,1] \big) \big( o(1) + \mathcal{O}(\nu) 
\big),$$
$0 < \nu \ll 1$. Since $\phi(r) \longrightarrow \infty$, then it follows 
easily from \eqref{eq+} that
\begin{equation}\label{eq1+}
\mathcal{N} \big( \mathcal{C}_{\alpha}(r,1) \cap \br \big)
= n \big( [r,1] \big) \big( 1 + o(1) \big) =
\phi(r) \big( 1 + o(1) \big), \quad r \searrow 0.
\end{equation}
Now, \eqref{eq6,13} together with the identities 
$\mathcal{N} \big( \mathcal{C}_{\alpha}(r,r_{0}) \cap \br \big) = 
\mathcal{N} \big( \mathcal{C}_{\alpha}(r,1) \cap \br \big) + \mathcal{O}(1)$
and  \eqref{eq1+} give (i).

\medskip

\noindent
(ii): If the magnetic field is constant, remember that we have $\zeta = 2b_0$. 
Thus, if the function $U$ satisfies $U \ge 0$, then \eqref{eq6,11} together 
with \eqref{eq6.18} imply that
\begin{equation}\label{eq1++}
n_+ \Big((2rb_{0})^{\frac{1}{2}},p_{0} U p_{0} \Big) \le 
n_+ \big(r,K_V(0) \big) \le n_{+} \big( 2rb_0,p_0 \textbf{\textup{W}} (I) p_0 \big), \quad r > 0.
\end{equation}
Recall that $\textbf{\textup{W}} (I) = \vert U \vert^2$ as function. Therefore,
if $U \ge 0$ satisfies Assumption (A), then \cite[Lemma 3.3]{rage} implies that 
the l.h.s. and the r.h.s. quantities of \eqref{eq1++} have the same first asymptotic
term as $r \searrow 0$. Namely as $r \searrow 0$, 
$n_+ \big((2rb_{0})^{\frac{1}{2}},p_{0} U p_{0} \big)
 = C_{m} (2b_0)^{-1/m} r^{-1/m} \big( 1 + o(1) \big)$ and 
 $n_{+} \big( 2rb_0,p_0 \textbf{\textup{W}} (I) p_0 \big) = 
 C_{m} (2b_0)^{-1/m} r^{-1/m} \big( 1 + o(1) \big)$, the constant $C_m > 0$ being 
defined by \eqref{eq2.11+}. This implies that
\begin{equation}
n_+ \big(r,K_V(0) \big)
 = C_{m} (2b_0)^{-1/m} r^{-1/m} \big( 1 + o(1) \big), \quad r \searrow 0.
\end{equation}
Then, (ii) follows from \eqref{eq6,13} together with \cite[Corollary 3.11]{bo}
and the identity $\mathcal{N} \big( \mathcal{C}_{\alpha}(r,r_{0}) \cap \br \big) = 
\mathcal{N} \big( \mathcal{C}_{\alpha}(r,1) \cap \br \big) + \mathcal{O}(1)$.



\begin{thebibliography}{99}


\bibitem{bo}
    \textsc{J.-F. Bony, V. Bruneau, G. Raikov},
    \textit{Counting function of characteristic values and magnetic resonances},
    Commun. PDE. \textbf{39} (2014), 274-305.


\bibitem{go}
     \textsc{I. Gohberg, E. I. Sigal},
     \textit{An operator generalization of the logarithmic residue theorem and Rouché's theorem},
     Mat. Sb. (N.S.) \textbf{84 (126)} (1971), 607-629. 
     
\bibitem{goh}
     \textsc{I. Gohberg, J. Leiterer},
     \textit{Holomorphic operator  functions of one variable  and applications},
     Operator Theory, Advances and Applications, vol. \textbf{192} Birkhäuser Verlag, 2009, Methods
     from complex analysis in several variables.    
    
\bibitem{ra}
    \textsc{G. Raikov},
    \textit{Spectral asymptotics for the perturbed 2D Pauli operator with oscillating magnetic fields. I. Non-zero mean value of the magnetic field},
     Markov Process. Related Fields \textbf{9} (2003), 775-794. 
     
\bibitem{rage}
    \textsc{G. Raikov},
    \textit{Low energy asymptotics of the spectral shift function for Pauli operators 
    with nonconstant magnetic fields}
    Publ. Res. Inst. Math. Sci. \textbf{46} (2010), 565–590.     
    
\bibitem{diom}
    \textsc{D. Sambou},
    \textit{Counting function of magnetic eigenvalues for non-definite sign perturbations},
    Operator Theory: Advances and Applications, \textbf{254} (2016), 205-221.

\bibitem{sim}
    \textsc{B. Simon},
    \textit{Trace ideals and their applications},
    Lond. Math. Soc. Lect. Not. Series, \textbf{35} (1979), Cambridge University Press.  
    

\end{thebibliography}
\end{document}